\documentclass{ws-ijmpd}
\usepackage[super,compress]{cite}
\begin{document}

\markboth{C.~R.~Arg\"uelles,R.~Ruffini}
{Are the most massive dark objects harbored ?}

\catchline{}{}{}{}{}
%
%%%%%%%%%%%%%%%%%%%%%%%%%%%%%%%%%%%%%%%%%%%%%%%%%%%%%%%%%%%%%%%%%%%%

\title{\textbf{ARE THE MOST SUPER-MASSIVE DARK COMPACT OBJECTS HARBORED AT THE CENTER OF DARK MATTER HALOS ?}}

\author{C.~R.~ARG\"UELLES\footnote{ICRANet, P.zza della Repubblica 10,Italy.}}

\address{ICRANet, P.zza della Repubblica 10\\
Pescara, I--65122 Pescara,
Italy\footnote{International Center for Relativistic Astrophysics Network, secretariat@icranet.org, Italy.}\\
carlos.arguelles@icranet.org}

\author{R.~RUFFINI}

\address{Dipartimento di Fisica and ICRA, Sapienza Universit\`a di Roma, P.le Aldo Moro 5\\
Rome, I--00185 Rome, Italy\\
ICRANet, P.zza della Repubblica 10\\
Pescara, I--65122 Pescara, Italy\\
ruffini@icra.it}

\maketitle

\begin{history}
\received{}
\revised{}
\end{history}

\begin{abstract}
\textit{ESSAY SELECTED FOR HONORABLE MENTION 2014 BY THE GRAVITY RESEARCH FOUNDATION}\\

We study an isothermal system of semi-degenerate self-gravitating fermions in General Relativity (GR). The most general solutions present mass density profiles with a central degenerate compact core governed by quantum statistics followed by an extended plateau, and ending in a power law behaviour $r^{-2}$. By fixing the fermion mass $m$ in the keV regime, the different solutions depending on the free parameters of the model: the degeneracy and temperature parameters at the center, are systematically constructed along the one-parameter sequences of equilibrium configurations up to the critical point, which is represented by the maximum in a central density ($\rho_0$) Vs. core mass ($M_c$) diagram. We show that for fully degenerate cores, the Oppenheimer-Volkoff (OV) mass limit $M_{c}^{cr}\propto M_{pl}^3/m^2$ is obtained, while instead for low degenerate cores, the critical core mass increases showing the temperature effects in a non linear way. The main result of this work is that when applying this theory to model the distribution of dark matter in big elliptical galaxies from miliparsec distance-scales up to $10^2$ Kpc, we do not find any critical core-halo configuration of self-gravitating fermions, able to explain both the most super-massive dark object at their center together with the DM halo simultaneously.

\end{abstract}

\keywords{Dark Matter; Galaxies: Super Massive Black Holes - Halos; Self-gravitating Systems: fermions}

\maketitle

\section{Introduction}

The nature and role of the dark matter (DM) particle at galactic scales together with the intricate relation involving the spatial distribution of its density, mass-to-light ratios and the estimation of the very massive dark object mass, are yet unsolved issues of major importance in physics and astrophysics.

Observationally, the presence of DM halos is safely confirmed in the smallest and less luminous ($10^3 L_\odot\lesssim L\lesssim 10^7 L_\odot$) galactic systems, i.e.: dwarf galaxies, where DM contributes $90\%$ or more to the total mass even much inside the half-light radius $R_e$. (see e.g. \cite{jpap1}). Instead, in the case of more luminous galaxies, both baryonic (i.e. stars) and dark matter contributes in comparable amounts to the total mass, being a big challenge to disentangle the gravitational effect of the DM component within $r\sim R_e$ (see e.g. \cite{edbk1} and refs. therein). In spiral galaxies the observation of extended HI regions in the disk structure provides an important and universal dynamical tracer which, through rotation curves analysis, has provided strong evidence for the existence of DM even up to several $R_e$ (see e.g. \cite{jpap2}).

However, in the case of big elliptical and early-type galaxies, most of them containing super massive dark objects at their centers (see e.g. \cite{jpap3}), there is no definitive evidence for the existence of DM halos. The low surface brightness beyond $R_e$ makes it a hard task to obtain reliable spectra to determine dispersion velocities. Among the several methods available to prove the mass distribution beyond $R_e$ in elliptical galaxies, such as integrated stellar light spectrum, globular cluster and planetary nebulae kinematics, diffuse X-ray observation or weak gravitational lensing; no evidence for DM halos has been found even out to few $R_e$ in many ellipticals by the use of kinematical methods (\cite{jpap4,jpap5}). Meanwhile, by using X-ray observations in an small sample of nearby ellipticals, as for example in \cite{jpap6}\, , a clear evidence for considerable amounts of DM at radii $r\sim10 R_e$ was given. In any case, the more interesting constraints on DM in early-type galaxies are restricted to the more massive systems, which are placed near the center of group or clusters. This implies to be a difficult task to confirm whether the existence of extensive halos are an inherent property of the galaxy itself, or whether corresponds to the group-scale (see e.g. \cite{edbk1} chapter $4.9.2$).

On theoretical and numerical grounds, the paradigm regarding the nature and spatial distribution of the DM particle in large and small distance-scales is centered in Newtonian N-body simulations within Lambda Cold Dark Matter ($\Lambda$CDM) cosmologies \cite{jpap7,jpap8}\, , being the beyond Standard Model (SM) elementary particle WIMP (Weakly Interacting Massive Particles) the preferred DM candidate. Nonetheless, despite the good agreement of this model with the large scale structure of the Universe, some subtle problems remains at galactic scales such as the core-cusp discrepancy and the lost satellites problem (see e.g. \cite{jpap9}).

An alternative and very promising beyond SM particle which has received increasing attention in cosmology and structure formation in the last decade, is the right handed sterile neutrino with masses of $\sim$ keV (see e.g. \cite{jpap10,jpap11,jpap12,jpap13})\,. Moreover, in \cite{jpap14} (see also refs. therein), Warm Dark Matter (WDM) halos has been obtained from numerical simulations solving the discrepancies which arises at galactic scales in the CDM paradigm, being again the sterile neutrino a plausible DM candidate.

Continuing on theoretical grounds and from a different and complementary perspective, the problem of modeling the distribution of dark matter in galaxies in terms of equilibrium configurations of collisionless self-gravitating fermionic particles has already been considered in \cite{jpap15,jpap16}\,. More recently in \cite{jpap17,jpap18,publ1,publ2,publ3}\,, this approach was developed in a fully relativistic treatment with applications to dark matter in normal galaxies being the spin $1/2$ fermion with masses $m\sim$few keV the preferred DM candidate in excellent agreement with halo observations (see e.g. \cite{publ2}). Again the sterile neutrino appears as an appealing candidate. An interesting characteristic in the density profile solutions of this kind of models with $m\sim10 keV$, is that they present a novel core-halo morphology composed by: i) a compact degenerate core of constant density at sub-parsec scales governed by quantum statistics (i.e. Pauli principle); ii) an intermediate region with a sharply decreasing density distribution followed by an extended plateau; iii) a decreasing $\rho\propto r^{-2}$ leading to flat rotation curves fulfilling classical Boltzmann statistics. In \cite{publ3}\,, and always for $m\sim 10 keV$, using observations from typical dwarf galaxies up to typical big spirals allowing for DM halo characteristics, it is clearly shown how the compact core described above in $i)$ may work as an alternative to intermediate ($M_c\sim10^4 M_\odot$) and super massive Black Holes (BH) (up to $M_c\sim10^7 M_\odot$) at their centers in simultaneous compatibility with each observed DM halo. It is then further shown in \cite{publ3} how, out of first principles, a possible universal correlation between the DM halos and the massive dark objects arises for $m=10 keV$. Interestingly, a very similar correlation law to the one theoretically found in \cite{publ3} in the range $M_c\in(10^6,10^7) M_\odot$ (with correspondent halo masses $M_h$ from $\sim 10^{11} M_\odot$ up to $\sim 10^{12} M_\odot$), has been found from observations in \cite{jpap19}\,, relating the dark halo masses to central mass concentrations, these last however identified by Ferrarese as black hole masses.

In this essay we present a detailed analysis of the above sketched model involving full Fermi-Dirac statistics, in an extreme relativistic regime which will allow us to deal with the most massive central dark objects of $M_c\sim 10^9 M_\odot$ at miliparsec scales. In this context, we present the density profiles and rotation curves for a fermion mass $m\sim 10 keV$, when the compact cores in all the solutions are very near Oppenheimer-Volkoff limit. Clearly, this approach makes the use of a General Relativistic treatment to be mandatory. In exploring the full range of the free model parameters such as temperature and degeneracy, we conclude that if a super massive dark object of $M_c\sim 10^9 M_\odot$ is formed at the center, no astrophysical DM halo structure should be present simultaneously in that system. A discussion to this respect in relation with observations and formation history is given in the third section.

\section{Theoretical Framework}

The system of Einstein equations are written in a spherically symmetric space-time metric $g_{\mu \nu}={\rm diag}(e^{\nu},-e^{\lambda},-r^2,-r^2\sin^2\theta)$,
where $\nu$ and $\lambda$ depend only on the radial coordinate $r$, together with the thermodynamic equilibrium conditions of Tolman \cite{jpap20}\ , and Klein \cite{jpap21}\, ,
\begin{equation}
e^{\nu/2} T=const.\, , \quad e^{\nu/2}(\mu+m c^2)=const, \nonumber
\end{equation}
where $T$ is the temperature, $\mu$ the chemical potential, $m$ the particle mass and $c$ the speed of light. We then write the system of Einstein equations in the following dimensionless way, following \cite{jpap16} and \cite{jpap18}\,,
\begin{align}
		&\frac{d\hat M}{d\hat r}=4\pi\hat r^2\hat\rho \label{eq:1}\\
		&\frac{d\theta}{d\hat r}=\frac{\beta_0(\theta-\theta_0)-1}{\beta_0}
    \frac{\hat M+4\pi\hat P\hat r^3}{\hat r^2(1-2\hat M/\hat r)}\\
    &\frac{d\nu}{d\hat r}=\frac{\hat M+4\pi\hat P\hat r^3}{\hat r^2(1-2\hat M/\hat r)} \\
    &\beta_0=\beta(r) e^{\frac{\nu(r)-\nu_0}{2}}\, . \label{eq:2}
\end{align}
The variables of the system are the mass $M$, the metric factor $\nu$ , the temperature parameter $\beta=k T/(m c^2)$ and the degeneracy parameter $\theta=\mu/(k T)$. The dimensionless quantities are defined as: $\hat r=r/\chi$, $\hat M=G M/(c^2\chi)$, $\hat\rho=G \chi^2\rho/c^2$ and $\hat P=G \chi^2 P/c^4$, with $\chi=2\pi^{3/2}(\hbar/mc)(M_p/m)$ and $M_p=\sqrt{\hbar c/G}$ the Planck mass.  The mass density $\rho$ and pressure $P$ are given by Fermi-Dirac statistics (with the particle helicity $g=2$):
\begin{align}
    \rho &= m\frac{2}{h^3}\int f(p)\left[1+\frac{\epsilon(p)}{m c^2}\right]\,d^3p,\label{eq:rho}\\
    P &= \frac13 \frac{2}{h^3}\int
    f(p)\left[1+\frac{\epsilon(p)}{m c^2}\right]^{-1}\left[1+\frac{\epsilon(p)}{2 m c^2}\right]\epsilon\,d^3p,\label{eq:p}
\end{align}
where the integration is extended over all the momentum space and $f(p)=(\exp[(\epsilon(p)-\mu)/(k T)]+1)^{-1}$. Here $\epsilon(p)=\sqrt{c^2 p^2+m^2 c^4}-mc^2$ is the particle kinetic energy, $\mu$ the chemical potential with the particle rest-energy subtracted off. We do not include the presence of anti-fermions, i.e.~we consider temperatures that always satisfy $T <m c^2/k$.

We want to further emphasize the central role of the Fermionic quantum satistics in the model, by recalling the necessity of the Pauli principle to form the central degenerate massive compact objects, mentioned above.

This system is solved for a fixed particle mass $m$ in the keV range, with initial conditions $M(0)=\nu(0)=0$, and given parameters $\theta_0>0$ (depending on the chosen central degeneracy), and $\beta_0$. We thus construct a sequence of different thermodynamic equilibrium configurations where each point in the sequence has different central temperatures $T_0$ and central chemical potential $\mu_0$, so that satisfy the $\theta_0$ fixed condition.

\begin{figure}[!hbtp]
\centering
\includegraphics[width=7.cm]{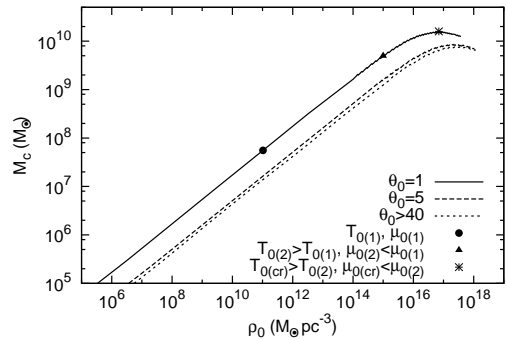}
\caption{Different sequences of equilibrium configurations plotted in a central density ($\rho_0$) Vs. core mass ($M_c$) diagram. The critical core mass is reached at the maximal value of $M_c$. Each sequence is built for selected values of $\theta_0=\mu_0/kT_0$ and different values of $T_0, \mu_0$ varying accordingly.}
\label{fig:1}
\end{figure}

Defining the core radius $r_c$ of each equilibrium system at the first maximum of its rotation curve, we represent the results obtained for each sequence in a central density ($\rho_0$) vs. core mass ($M_c$) diagram (see Fig.\ref{fig:1}). It is shown that the critical core mass $M_c^{cr}$ is reached at the maximum of each $M_c(\rho_0)$ curve.

It is important to emphasize that the method followed here to define the critical points along each family of thermodynamic equilibrium configurations, fulfills the \textit{turning point} definition given in \cite{jpap22}\,, which allows them to formally demonstrate Sorkin's theorem \cite{jpap23} showing the existence of a thermodynamic instability on one side of the turning point
\footnote{In reality, to properly implement the formal concept of \textit{turning point} as used for example in \cite{jpap22}\,, the \textit{total} mass of the system $M_t$ (previous choice of a cut-off in the momentum space at i.e. $r_t\sim10 r_h$ to define it) should be used in the central density ($\rho_0$) Vs. mass ($M$) diagram, instead of the core mass $M_c$. Nonetheless, considering that in fully degenerate systems the critical masses $M_c^{cr}$ are basically equal to the OV mass, it should imply that the extended and diluted halo plays no relevant role in the physics near the critical point, in some analogy to the case of Super-Nova core collapse where only the fully degenerate core is considered in the process.}.

In Table I we show a set of central critical parameters of the model together with the correspondent critical core masses, for a very wide range of fixed central degeneracy parameters $\theta_0$ and particle mass $m=8.5$ keV$/c^2$.

\begin{table}[h!]
\tbl{Critical temperature parameter and normalized chemical potential at the center of each different critical configuration, for different fixed central degeneracies.}
{\begin{tabular}{@{}cccc@{}} \toprule
$\theta_0$ & $\beta_0^{cr}$ & $\mu_0^{cr}/mc^2$ & $M_c^{cr} (M_\odot)$ \\
\colrule
1 & $6.45\times10^{-2}$ & $6.45\times10^{-2}$ & $1.59\times10^{10}$ \\
 5 & $2.23\times10^{-2}$ & $1.11\times10^{-1}$ & $7.91\times10^9$ \\
 40 & $8.33\times10^{-3}$ & $3.33\times10^{-1}$ & $7.44\times10^9$ \\
 55 & $6.06\times10^{-3}$ & $3.33\times10^{-1}$ & $7.44\times10^9$ \\
 100 & $3.33\times10^{-3}$ & $3.33\times10^{-1}$ & $7.44\times10^9$ \\ \botrule
\end{tabular} \label{table:1}}
\end{table}

Defining the halo radius of each configuration at the onset of the flat rotation curve, we show in Table II the critical halo magnitudes $r_h^{cr}$, $M_h^{cr}$ and $v_h^{cr}$ corresponding to the same set of critical parameters as given in Table I.

\begin{table}[h!]
\tbl{Critical halo magnitudes of different critical configurations, for different fixed central degeneracies as given in Table I.}
{\begin{tabular}{@{}cccc@{}} \toprule
$\theta_0$ & $r_h^{cr}$ & $M_h^{cr}/mc^2$ & $v_h^{cr}$ \\
& (pc) & ($M_\odot$) & (km/s) \\ \colrule
1 & $4.4\times10^{-1}$ & $5.7\times10^{11}$ & $7.5\times10^4$ \\
5 & $4.0\times10^{-1}$ & $4.3\times10^{11}$ & $6.2\times10^4$ \\
40 & $4.3\times10^{3}$ & $1.1\times10^{15}$ & $3.3\times10^4$ \\
55 & $2.9\times10^{5}$ & $6.0\times10^{16}$ & $2.9\times10^4$ \\
100 & $2.0\times10^{11}$ & $2.3\times10^{22}$ & $2.2\times10^4$ \\ \botrule
\end{tabular} \label{table:2}}
\end{table}

The results obtained in Tables I and II imply a marked division in two different families depending on the value of $M_c^{cr}$.

\textit{i}) The first family: the critical mass has roughly a constant value $M_c^{cr}=7.44\times10^9 M_\odot$. This family corresponds to large values of the central degeneracy ($\theta_0\geq40$), where the critical temperature parameter is always lower than $\beta_0^{cr}\lesssim8\times10^{-3}$ and the critical chemical potential $\mu_0^{cr}\approx$ const. Physically, these highly degenerate cores are entirely supported against gravitational collapse by the degeneracy pressure. In this case the critical core mass is uniquely determined by the particle mass according the relation $M_c^{cr}\propto M_{pl}^3/m^2$ (see also last section).

\textit{ii}) The second family: the critical core mass increases from $M_c^{cr}=7.44\times10^9 M_\odot$ up to $M_c^{cr}\sim10^{10} M_\odot$. This case corresponds to critical cores with a lower central degeneracy compared with the former family ($1<\theta_0<40$). Here the critical temperature parameter ($\beta_0\sim10^{-2}$), is closer to the relativistic regime with respect to the first family. This result physically indicates that the thermal pressure term has now an appreciable contribution to the total pressure, which supports the critical core against gravitational collapse. In this case $M_c^{cr}$ is completely determined by the particle mass $m$, the central temperature $T_0^{cr}$ and the central chemical potential $\mu_0^{cr}$ (see last section).

\begin{figure}
 \centering
\includegraphics[width=7.cm]{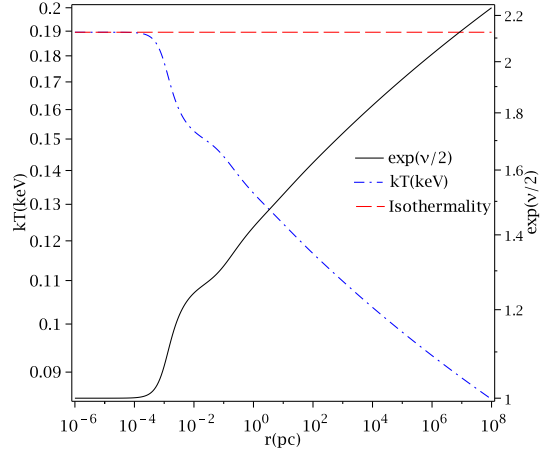}
\caption{The critical temperature profile of the system (in keV) and the critical metric, for $\theta_0=5$ and $\beta_0^{cr}=2.23\times 10^{-2}$. The dashed line corresponds to the isothermality condition, $Te^{\nu/2}=const$.}
\label{fig:2}
\end{figure}

\begin{figure}
 \centering
\includegraphics[width=7.cm]{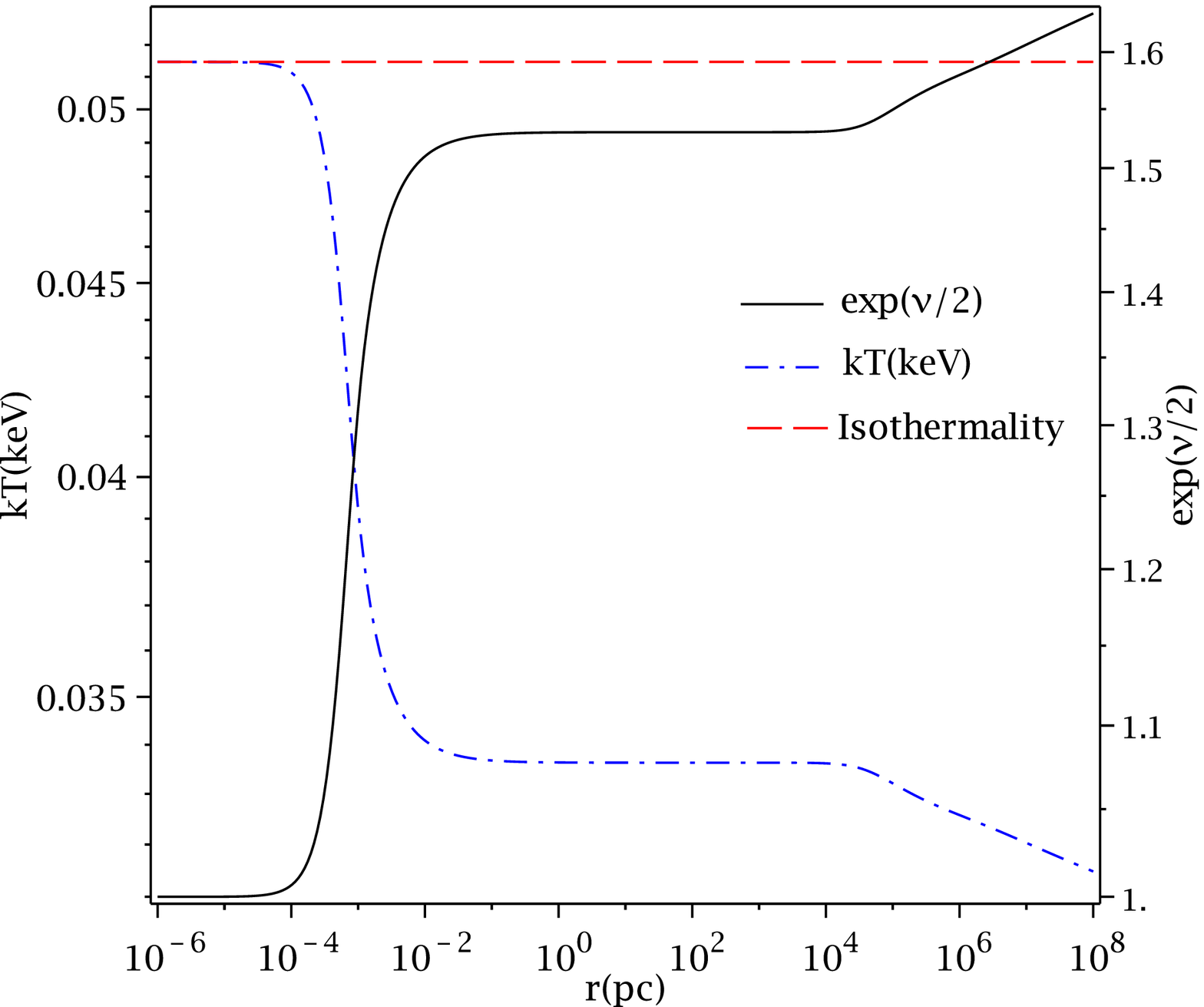}
\caption{The critical temperature of the system (in keV) and the critical metric, for $\theta_0=55$ and $\beta_0^{cr}=6.06\times10^{-3}$. The dashed line corresponds to the isothermality condition, $Te^{\nu/2}=const$.}
\label{fig:3}
\end{figure}

In Figs.~(\ref{fig:2}) and (\ref{fig:3}) we show a critical metric factor $e^{\nu/2}$ and a critical temperature $kT$ as a function of the radius for the two different families mentioned above.

%%%%%%%%%%%%%%%%%%%%%%%%%%%%%%%%%%%%%%%%%%%%%%%%%%%%%%%%%%%%%%%%%%%%%%%%%%%%%
\section{Astrophysical Application and Discussion}
We will now attempt to use the critical configurations obtained before to explain the DM distribution in big galactic halos, as well as providing an alternative candidate to the standard central black hole paradigm. From now on, we will use a fixed particle mass of $m=$8.5 keV$/c^2$, being this choice motivated by the fact we want to deal with super massive dark objects having critical core masses of the order $M_c^{cr}\propto m_{pl}^3/m^2\sim10^9 M_{\odot}$. Moreover, such a relativistic object would have an OV radius $R_{OV}$ very near the Schwarschild radius $R_s$ ($R_{OV}\sim 3 R_s$), and then practically indistinguishable from a BH of the same mass.

\begin{figure}[!hbtp]
\centering
\includegraphics[width=7.cm]{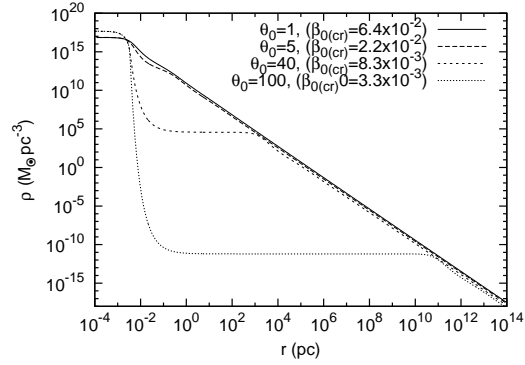}
\caption{Critical density profiles for different values of $\theta_0$ with the correspondent critical temperature parameters $\beta_0^{cr}$.}
\label{fig:4}
\end{figure}

\begin{figure}[!h]
\centering
\includegraphics[width=7.cm]{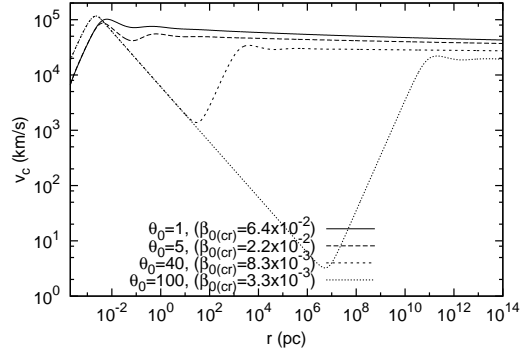}
\caption{Critical rotation curves for different values of $\theta_0$ as given in Fig.~\ref{fig:4}. To note the high values of $v_c(r)\sim10^4$ km/s in the flat parts of each curve due to the high critical temperature parameters $\beta_0^{cr}$.}
\label{fig:5}
\end{figure}

\begin{figure}[!h]
\centering
\includegraphics[width=7.cm]{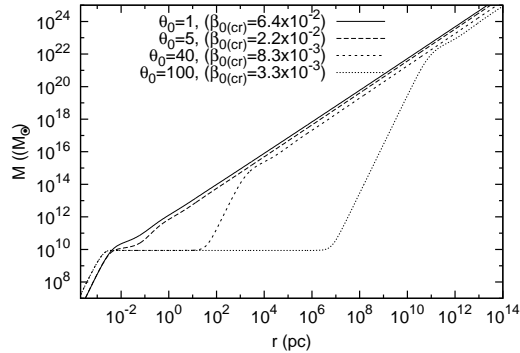}
\caption{Critical mass profiles for different values of $\theta_0$ as given in Figs.~\ref{fig:4}--\ref{fig:5}.}
\label{fig:6}
\end{figure}

In Figs.~(\ref{fig:4}), (\ref{fig:5}) and (\ref{fig:6}) we show different critical density profiles, critical rotation curves and critical mass profiles respectively, for a wide range of different central degeneracy parameters.

From Fig.~\ref{fig:5} and Table II we see that critical configuration of self-gravitating fermions with central degeneracies ranging from $\theta_0=1$ (i.e. mainly thermal pressure supported cores) up to $\theta_0=100$ (i.e. mainly degeneracy pressure supported cores), have flat rotation velocities $v_h^{cr}\sim10^4$ km/s. Even if for $\theta_0\sim50$ the halo sizes could match observations, the halo masses are well above any observed value and so the circular velocities; further implying that none of these mathematical solutions are compatible with any observable astrophysical systems. However, since larger values of $\theta_0$ imply lower values for $\beta_0^{cr}$, as shown in Table I, there is a point (at $\theta_0\sim 10^6$ and so $\beta_0^{cr}\sim10^{-7}$) where the \textit{halo} rotation curves reach the typical observed values as in spiral or elliptical galaxies of $v_h^{cr}\sim10^2$ km/s \footnote{These values of rotation curves were used in the very low (spetial) relativistic regime version of this model (i.e. $\theta_0\sim10^1$), as presented in \cite{publ1,publ2,publ3}\,, and in that case leading to the correct halo masses and sizes in galaxies.}. Nonetheless, since already at $\theta_0\gtrsim80$ (see for example $\theta_0=100$ in Fig.~\ref{fig:4}) the plateau region of the density profile is lower than the mean DM density of the Universe ($\rho_{uni}\sim3H_0^2/(4\pi G)$), then higher central degeneracies will imply even more diluted halos (i.e. already disappeared), never reaching typical flat rotation curves at $10^{1-2}$ Kpc halo distance-scales. Thus, the critical configurations here analyzed in the fully degenerate regime $\theta_0\gg1$, should only consist on a central super-massive compact objects of $M_c^{cr}\sim10^9 M_\odot$ consisting on DM particles of $m\sim 10$keV.

In the light of the present analysis we then conclude that there is no critical core-halo configuration of self-gravitating DM fermions, able to explain both the most super-massive dark object at the center together with an outer DM halo simultaneously.

The concept of simultaneous co-existence of super-massive dark objects and DM halos at some cosmological epoch $z$ is of central importance to better understand the structure growth, galaxy formation history and evolution. An important observational result aiming in these directions has been reported in \cite{jpap24}\, . In that work there is clear evidence for evolution of early-type galaxies evolving from $z\sim 2.5$ (consistently studied only in terms of light profiles) up to now ($z\sim0$) enlarging their sizes and masses (lowering in density), which imply necessarily subsequent gathering of matter from larger-scale environments in their complex evolutionary history (probably in the form of dark and/or baryionic matter). Thus, contrasting the results obtained in this work with observational results for big ellipticals as the ones already recalled in \cite{jpap24}\, and \cite{jpap4,jpap5,jpap6}\,, but also in \cite{jpap25,jpap26} for the more relevant case of the giant elliptical M87 with the detected super-massive dark object of $M_c\sim 7\times10^9 M_\odot$; could well imply that even if there is clear evidence in some cases of co-existence of central dark massive objects and dark halos at $z\sim0$, this could not be the case at early stages i.e.$z\sim 3$. The connection between the formation of super-massive compact dark objects at early epochs, the relation with the host halo and possible subsequent accretion form their larger-scale environments are still important open questions.

%%%%%%%%%%%%%%%%%%%%%%%%%%%%%%%%%%%%%%%%
%-----------------------------------------
 \section{An Analytical Expression for the Critical Mass}
What one may find very insightful is to derive an analytical (approximate) formula for the critical mass, so that one can then adventure to explore and understand the subjacent physics of it. For this we will
use the Newtonian hydrostatic equilibrium equation corresponding to the last stable configuration, where the pressure due to gravity is balanced by a high relativistic semi-degenerate Fermi gas :
{\small
\begin{eqnarray}
P_g(r)&=&P_T^{ur}(r), \nonumber \\
\frac{GM(r)\rho(r)}{r}&\approx&\frac{\mu^4}{12\pi^2(\hbar c)^3}+\frac{\mu^2(kT)^2}{6\sqrt\pi(\hbar c)^3},
\label{lastequil}
\end{eqnarray}
}
where $P_T^{ur}(r)$ is the ultra relativistic approximation of a highly relativistic Fermi gas ($\mu\gg mc^2$), which has been expanded up to second order in temperature around $\mu/kT\gg1$ (see e.g. \cite{edbk2}). We have used in \ref{lastequil} the fermi energy ($\epsilon_f=\mu$) with the rest energy substracted-off in consistency with the theoretical formulation of our model.

Then, after constant density considerations for $r\lesssim r_c$ in the ultra-relativistic approximation here adopted, we get,
\begin{equation}
M_c^{cr}\approx\frac{3\sqrt{\pi}}{16}\frac{M_{pl}^3}{m^2}\left(1+\frac{2\pi^2}{\theta_0^2}\right)^{3/2}.
\label{Mcranalyt}
\end{equation}

It is clear from this equation that for high central degenerate systems ($\theta_0\gg\sqrt{2}\pi$), the critical core mass $M_c^{cr}$ is independent of $\theta_0$ and then proportional to $M_{pl}^3/m^2$. However, for low values of the central degeneracy ($\theta_0\sim\sqrt{2}\pi$) the second term in (\ref{Mcranalyt}) starts to be relevant, showing the finite temperature effects. In fact, using $\theta_0=40$, we have $M_c^{cr}=7.62\times 10^9\,M_{\odot}$, just a 2\% difference with the numerical result of $7.44\times 10^9\,M_{\odot}$. However, for $\theta_0=5$ we have $M_c^{cr}=1.79\times 10^{10}\,M_{\odot}$, almost a factor 2 above the numerical value of $7.91\times 10^9\,M_{\odot}$ obtained in Table I. This shows that our approximation of an ultra-relativistic fermi gas in newtonian equilibrium breaks down and a fully relativistic treatment is needed.

%%%%%%%%%%%%%%%%%%%%%%%%%%%%%%%%%%%%%%%%%%%%%%%%%%%%%%%%%%%%%%

The authors wish to thank Prof. Herman Mosquera Cuesta for critical reading of the manuscript. The authors also want to thank Dr. Jorge A. Rueda for valuable suggestions on some technical points in the application of the model in the actual work.

%%%%%%%%%%%%%%%%%%%%%%%%%%%%%%%%%%%%%%%%%%%%%%%%%%%%%%%%%%%%%%
%%									 References                         %%
%%%%%%%%%%%%%%%%%%%%%%%%%%%%%%%%%%%%%%%%%%%%%%%%%%%%%%%%%%%%%%

%\bibliographystyle{apsrev}

%%%%%%%%%%%%%%%%%%%%%%%%%%%%%%%%%%%%%%%%%%%%%%%%%%%%%%%%%%%%%%

\end{document}